\documentclass[12pt,a4paper]{JHEP}

\title{S-Duality and Brane Descent Relations}
\author{Laurent Houart\\Theoretical Physics Group\\
Blackett Laboratory,
Imperial College \\
London SW7 2BZ, UK\\
\email l.houart@ic.ac.uk}
\author{Yolanda Lozano\\Theory Division, CERN\\
1211 Gen\`eve 23, Switzerland\\
\email yolanda.lozano@cern.ch}
\abstract{We present a description of the type IIB
NS-NS p-branes in terms of topological solitons in 
systems of spacetime-filling brane-antibrane pairs.
S-duality implies that these spacetime-filling branes are
NS9-branes, S-dual to the D9-branes of the type IIB theory.
The possible vortex-like solutions in an NS9, anti-NS9
configuration are identified by looking at its worldvolume
effective action. Finally we discuss the implications of these 
constructions in the
description of BPS and non-BPS states in the strongly coupled 
Heterotic SO(32) theory.}

\preprint{CERN-TH/99-354\\
Imperial/TP/99-0/13}

\makeatletter
\@addtoreset{equation}{section}
\makeatother

\begin{document}

\section{Introduction}

Recently, very significant progress has been made in our understanding  
of the role of non-BPS configurations in String Theory, and interesting 
connections between some BPS branes and non-BPS branes have been 
uncovered (see \cite{reva} for reviews and references therein).
It is by now well-known that the BPS Dp-branes of the type II theories can
be understood as bound states of a certain number of spacetime-filling
branes or brane-antibrane pairs in the theory. 
In the type IIB case one considers spacetime-filling brane-antibrane
pairs built out of BPS type IIB D9-branes charged with respect 
to the RR 10-form potential \cite{Witten}, whereas in type IIA the
spacetime-filling branes are 
the non-BPS D9-branes introduced by Ho\u{r}ava in \cite{Horava}.
The BPS Dp-branes appear as topological defects in the worldvolume
of these unstable systems. An improved understanding
of the physics of tachyonic excitations present in unstable 
branes or brane-antibrane pairs has been the key ingredient 
for these developments \cite{Sen0}. The presence of 
tachyons is no longer viewed as a severe pathology of the theory,
and a tachyon can be considered like a true Higgs field 
which tends to develop a stable vacuum expectation value. 
The simplest example is the
description of a BPS Dp-brane in terms of the condensation of the 
tachyonic mode of the open string stretching between a pair of D(p+2)
anti-D(p+2) branes \cite{Sen1}, process which can be iterated to
construct lower dimensional D-branes by brane descent relations.

Starting with an unstable system of spacetime-filling 
brane-antibrane pairs, 
one can analyse the homotopy groups of the 
vacuum manifold of the Higgs field (tachyon) which appears in the system,
and deduce the possible stable defects corresponding to 
stable D-branes \cite{Witten,Horava,OS}. In the type IIB case, the
system of coincident D-branes and anti D-branes is
in fact characterised by  equivalence classes of pairs of vector bundles
(one for the D-branes and one for the anti D-branes) which correspond to 
configurations of spacetime-filling branes up to creation and
annihilation, and permit to classify the stable D$p$-brane charges
\footnote{In type IIA, since one starts with unstable uncharged D9-branes,
describing the conserved charges in terms of equivalence classes is less
direct \cite{Horava}.}. These classes appear to be described in a 
mathematical framework in terms of K-theory groups \cite{MM,Witten}.
More precisely, in the type IIB case the stable D$p$-brane charges
are classified by the K-theory groups $\tilde K (S^{9-p})$, where 
the sphere $S^{9-p}$ represents the dimensions transverse to the
worldvolume of the $p$-brane (compactified by adding a point at infinity)
and $\tilde K$ stands for the subgroup of $K$ defined by restricting
the equivalence classes to the same number of branes and antibranes.
There is a close relation between K-theory (which characterises the
conserved D-brane charges and configurations of spacetime-filling branes
up to creation and annihilation) and homotopy theory (which classifies
the topological defects). The K-theory groups of spheres are equal to
the homotopy groups of the vacuum manifold of the Higgs field that
appears in the worldvolume theory of the system of spacetime-filling
D9 anti-D9 branes. 

These beautiful developments in our understanding of BPS and non-BPS
states refer mainly to D-branes, and very little is 
known about possible extensions to the NS-NS charged
p-branes (see however \cite{Yi,LY} for some qualitative results).
In this paper, we propose to make a step further in 
this direction and study in the framework of type IIB 
theory such possibilities.
We consider the S-dual picture of the usual spacetime-filling 
brane-antibrane construction of type IIB, and by studying a system 
of coincident NS9 anti-NS9
branes discuss the different possible topological defects in
the type IIB and  Heterotic SO(32) theories.  

The paper is organised as follows. In section 2 we review the
construction of the BPS Dp-branes of type IIB as bound states of
spacetime-filling D9 anti-D9 pairs. Section 3 discusses the S-dual
picture and shows how the NS-NS branes of the type IIB theory
can be understood as bound states of NS9 anti-NS9 systems.
We then turn in section 4 to the study of the  p-branes of the
Heterotic SO(32) theory as bound states of NS9 anti-NS9 branes in
a Heterotic background. In particular, we consider the description
of the Heterotic SO(32) theory as an orientifold of strongly coupled
type IIB by the S-dual of worldsheet parity reversal 
\cite{Hull2, BEHHLvdS}.
Finally, section 5 contains the conclusions.

\section{Type IIB Dp-branes as bound states}

In this section we briefly review the construction of the BPS Dp-branes
of the type IIB theory as bound states of a certain number of
spacetime-filling D9 anti-D9 pairs \cite{Witten}.
The BPS Dp-branes appear as topological solitons in the worldvolume
of this system, and the explicit way in which they couple
can be simply read from the worldvolume effective action of the 
9, ${\bar 9}$ pairs. 

The Wess-Zumino part of the worldvolume effective 
action of a Dp-brane is given by \cite{Doug}:

\begin{equation}
\label{uwz}
S= \int_{\Sigma_{p+1}} C \wedge {\rm Tr} e^{F}
\end{equation}

\noindent where C is the complex of differential forms which is the
sum of the RR potentials. More specifically, in the case
of N coincident D9-branes we are interested in, one has:

\begin{eqnarray}
\label{fwz}
S_{{\rm D9}}^{{\rm WZ}}&=&\int_{R^{9+1}} {\rm Tr}
\left[ C^{(10)}+C^{(8)}\wedge F+
C^{(6)}\wedge F\wedge F+C^{(4)}\wedge F\wedge F\wedge F+\right. 
\nonumber\\
&&\left. +C^{(2)}\wedge F\wedge F\wedge F\wedge F+
C^{(0)}F\wedge F\wedge F\wedge
F\wedge F\right]\, ,
\end{eqnarray}

\noindent where $C^{(p)}$ denotes a $p$-form RR-potential,
$F=db^{(1)}\in$ U(N) \footnote{We do not include explicitly the coupling
to the NS-NS 2-form in the field strength of the vector field since it
will be irrelevant in the discussion that follows.}
represents the
flux of a fundamental string ending on the branes, and the trace is
taken over the U(N) indices.

The effective action describing brane-antibrane pairs
contains additional worldvolume fields. One has, in addition to $F$, 
a second field strength $F^{\prime}$, associated with the antibranes,
and a complex charged tachyon field $T$. 
The WZ term has a similar but slightly
more involved structure \cite{KW}. It contains a term :
\begin{equation}
\label{pwz}
 \int  C \wedge  (e^F-e^{F^{\prime}})  
\end{equation}
along with tachyon-dependent terms which we will not consider here since 
they will not be important for our discussion. In fact, to discuss
the possible realisations of D-branes as topological defects 
one can focalise on the WZ terms of the brane action (\ref{fwz}) instead
of the ones of the brane-antibrane pairs (\ref{pwz}), bearing in mind
that the topologically non-trivial character of the soliton can be 
carried by one of the two field strengths, say $F$ 
(see for instance \cite{Sen1}).

We now turn to the analysis of the WZ terms of (\ref{fwz}).
The second term in this action provides a description of the D7-brane
as a bound state D9, anti-D9. The D9, anti-D9 system is unstable due
to the existence of a tachyonic mode in the open strings connecting
the two types of branes. The tachyon can condense in a topologically
non-trivial way \cite{Sen1}, and this condensation is accompanied by a 
localised magnetic flux such that:

\begin{equation}
\int_{R^2} {\rm Tr} F={{\rm integer}}\, .
\end{equation}

\noindent Therefore a vortex soliton
behaving as a D7-brane is induced, 
since the integration of ${\rm Tr}F$ gives
a coupling $\int_{R^{7+1}}C^{(8)}$ in the D9, anti-D9 
effective action.

Similarly, the other terms in this action describe other
type IIB D$p$-branes as bound states of a certain number of pairs 
D9, anti-D9. Generically, a $p$-brane can be realised, 
by a stepwise reasoning,  
as a bound state of a system of $n$ ($p+2k$)-brane anti-brane pairs for
$n=2^{k-1}$ \cite{Witten}, since the homotopy
group of the U($n$) vacuum manifold of the $n$ brane-antibrane 
pairs\footnote{The gauge group of the $n$ brane-antibrane pairs is
U($n$)$\times$ U($n$), but it is broken in the vacuum to the diagonal 
U($n$) by the tachyon field, and the vacuum manifold of the tachyon is
U($n$)$\times$U($n$)/U($n$), which is topologically equivalent to 
U($n$).}, which classifies the existence of
topologically non-trivial gauge fields on $R^{2k}$, satisfies:
$\Pi_{2k-1}$(U($n$))$=Z$.
Let us consider for instance the term:

\begin{equation}
\label{Cseis}
\int_{R^{9+1}}{\rm Tr}\left[C^{(6)}\wedge F\wedge F\right]
\end{equation}

\noindent in the D9, anti-D9 system effective action. 
$\int_{R^4}{\rm Tr} F\wedge F$ gives an
integer for $F\in U(2)$, since for
$k=2$ $n=2$. The structure of the WZ term implies that this 
instanton-like configuration carries $C^{(6)}$ charge, and can
therefore be identified as a D5-brane.
In this way the D5-brane is realised as a bound state of 2 D9, 
anti-D9 pairs of branes. The structure of the rest of the terms in
the WZ action completes a general pattern:
D7=(D9, anti-D9), D5=2 (D9, anti-D9), D3=4 (D9, anti-D9),  
D1=8 (D9, anti-D9) and D(-1)=16 (D9, anti-D9), of type IIB D$p$-branes as
bound states of D9, anti-D9 pairs. This realisation of the
BPS D$p$-branes as topological defects is coherent with the fact
that the reduced K-theory groups classifing the conserved D-brane
charges in type IIB are given by \cite{Witten} $\tilde K (S^n)$, with
$n=9-p$, and that these groups are related to the homotopy groups 
(see for instance \cite{OS}) as: $\tilde K (S^n)  = \pi_{n-1}(U(N))$ 
for sufficiently large $N$ (i.e. in the stable range). 
One thus has $\tilde K (S^n) = Z$ for $p$ odd and  
$\tilde K (S^n) = 0$ for $p$ even.

\section{The S-dual picture}

One important property of the type IIB theory is however missing from
the previous construction, namely its S-duality symmetry,
also manifest at the level of the $p$-brane solutions.
This symmetry implies that if, in particular, the D1- and the D5-branes 
can be understood as vortex-like solutions in pairs of D9, anti-D9
branes, the same should also be possible for  
the F1- and NS5-branes, in this case as topological solitons in
systems of NS9, anti-NS9 branes. 

The so-called NS9-brane is by definition
the S-dual of the D9-brane of the type IIB theory, and it is charged
with respect to a non-dynamical 10-form NS-NS field, S-dual to the 
$C^{(10)}$ RR-potential to which the D9-brane is minimally coupled 
\cite{tomasetal}.
This brane is indeed predicted by the analysis of the type IIB
spacetime supersymmetry algebra \cite{Hull1}. Its effective action 
was constructed in \cite{BEHHLvdS} by performing an S-duality
transformation in the D9-brane worldvolume
effective action. In this reference, and previously in \cite{Hull2}, 
it was also argued that the NS9-brane should play a role 
in the construction of the
Heterotic string with gauge group SO(32) as a nonperturbative 
orientifold of the type IIB
theory. We will comment on the details of this
construction in the next section.

The NS9-brane effective action
predicts the F1- and the NS5-branes as bound states of 8 and 2, 
respectively, NS9 anti-NS9 pairs of branes.
The analysis is the same than the one we have just made from the
D9-brane worldvolume effective action.
The D3-brane, being a singlet under
the SL(2,Z) S-duality group, can be realised as a bound state of either
D9, anti-D9 or NS9, anti-NS9 branes, since both types of branes carry
RR 4-form charge. The same thing happens with the 7-brane and the 
instanton, though in this case the explicit form of the solutions
and worldvolume effective actions describing the branes at strong
coupling is different from that in the weak coupling regime
(see \cite{EL}).
In all these configurations
the object whose tachyonic mode condenses is a D1-brane, stretched
between the NS9, anti-NS9 pairs.  
The explicit form of the NS9-brane worldvolume effective action reads
\cite{BEHHLvdS}:

\begin{eqnarray}
\label{NS9ac}
&&S_{{\rm NS9}}^{{\rm WZ}}=\int_{R^{9+1}}{\rm Tr}\left[ B^{(10)}+
{\tilde C}^{(8)}\wedge {\tilde F}+B^{(6)}\wedge {\tilde F}
\wedge {\tilde F}+
C^{(4)}\wedge {\tilde F}\wedge {\tilde F}\wedge {\tilde F}
+\right. \nonumber\\
&&\left. +B^{(2)}\wedge {\tilde F}\wedge {\tilde F}\wedge {\tilde F}
\wedge {\tilde F}+\frac{C^{(0)}}{(C^{(0)})^2+e^{-2\phi}}
{\tilde F}\wedge {\tilde F}\wedge {\tilde F}
\wedge {\tilde F}\wedge {\tilde F}\right]\, .
\end{eqnarray}

\noindent Here ${\tilde F}$ describes the
flux of a D1-brane ending on the NS9-branes\footnote{We have again
omitted the contribution from the RR 2-form potential.}, 
$B^{(10)}$ is the NS-NS
10-form potential with respect to which the NS9-brane is charged and
${\tilde C}^{(8)}$ is the S-dual of the 8-form RR-potential 
(see \cite{EL}). The second term in this
action describes a 7-brane, realised as a bound state of one pair of
NS9, anti-NS9 branes. This configuration was predicted in
\cite{LY}, as implied by the realisation of the M-theory Kaluza-Klein
monopole as a bound state of a pair M9, anti-M9.
The third term in (\ref{NS9ac}) is the S-dual of the term: 
$\int_{R^{9+1}}{\rm Tr}[C^{(6)}\wedge F\wedge F]$, that we
discussed in the previous section,
and describes a NS5-brane as a bound state
of two pairs of NS9, anti-NS9 branes. Similarly, the $B^{(2)}$-term 
describes a fundamental string as a bound state of eight pairs of
NS9, anti-NS9 branes.

The structure of the NS9-brane WZ terms implies that the NS-NS branes
of the type IIB theory can be understood as bound states of
NS9, anti-NS9 pairs of branes. The general
pattern that is derived from this analysis is: D7=(NS9, anti-NS9),
NS5=2 (NS9, anti-NS9), D3=4 (NS9, anti-NS9), F1=8 (NS9, anti-NS9)
and D(-1)=16 (NS9, anti-NS9), of type IIB p-branes as bound states 
of NS9, anti-NS9 pairs. 
Therefore, we see that all the branes predicted by the analysis of
the type IIB spacetime supersymmetry algebra, apart from the pp-wave
and the Kaluza-Klein monopole, can be realised as
bound states of any of the two types of spacetime-filling branes of
the theory. Of course for this to hold the NS9-branes
must be considered on an equal footing with the RR 9-branes.   
The pp-wave and the Kaluza-Klein monopole solutions are only defined
in spacetimes with one special, isometric, direction.
In the case of the pp-wave this is the direction of propagation
of the wave, whereas in the monopole case this is the 
Taub-NUT fiber of the transverse space. Therefore these branes 
cannot be understood as bound states of spacetime-filling branes 
that do not see any of these special directions.

\section{Heterotic SO(32) branes as bound states}

In this section we discuss how the solitonic branes of the Heterotic
SO(32) theory can be understood as bound states of spacetime-filling
branes. We start by reviewing the type I case, since our description
will strongly rely on the S-duality connection between the two theories.

\subsection{The Type I case}

All stable branes in the type I theory, including also non-BPS ones,
can be realised as bound states of a certain number of D9, anti-D9
pairs of branes \cite{Witten}.
The definition of the type I theory as an orientifold of
type IIB by its worldsheet parity reversal symmetry makes clear 
that the analysis of section 2  holds straightforwardly 
in this case,
with the only consideration of the projections induced by the orientifold
construction in the different gauge fields. In particular the U(N)
gauge group of a set of N coincident D-branes is broken to SO(N) for
D1- and D9-branes and to Sp(N) for D5-branes \cite{Witten2}.

The general pattern of
stable\footnote{The D7 and D8 non-BPS branes are however unstable due
to the presence of a
tachyonic mode in the open strings with one end on the brane and the
other on one of the 32 D9-branes \cite{FGLS}.}
type I branes in terms of bound states of D9, anti-D9 pairs has
been derived in \cite{Witten}, and it is given by:
D8=(D9, anti-D9), D7=2 (D9, anti-D9), D5=4 (D9, anti-D9),
D1=8 (D9, anti-D9), D0=16 (D9, anti-D9), D(-1)=32 (D9, anti-D9).
The D8, D7, D0 and D(-1) branes are non-BPS.
In the type I theory some stable\footnote{The brane can end up however being
unstable due to the presence of tachyonic modes in the open strings with
one end on the 32 D9-branes. This happens to be the case for the 
D7 and D8 branes, as we have just mentioned.}, non-BPS, 
p-branes with p even can be realised as 
bound states of BPS (p+1), anti-(p+1) branes, because the tachyonic
mode that survives in the spectrum in the type IIB case can be projected
out by the orientifold construction for certain values of p
\cite{Sen1}, \cite{Sen2}.
In particular the D8 can be obtained as
D8=(D9, anti-D9) and the D0 as D0=(D1, anti-D1)=16 (D9, anti-D9). The
D7 and D-instanton are obtained as: D7=(D7, anti-D7), 
D(-1)=(D(-1), anti-D(-1)) from type IIB branes, since in these cases
the orientifold
projection exchanges the brane with the antibrane and eliminates
the tachyon. Therefore,
D7=(D7, anti-D7)=2 (D9, anti-D9), D(-1)=(D(-1),anti-D(-1))=32 (D9,
anti-D9), using that in type IIB: D7=(D9, anti-D9), 
D(-1)=16 (D9, anti-D9) \cite{Witten}.

The general pattern of stable type I branes shows
in particular that a same BPS brane in the
type I and type IIB theories can be obtained as a bound state of a 
different number of spacetime-filling branes and antibranes.
A concrete example of how the number of spacetime-filling branes required in
the construction changes from
the type IIB to the type I theory 
is the D5-brane, discussed in \cite{Witten}.
In the previous section we showed how $\Pi_3 (U(2))=Z$ implied that
the type IIB D5-brane could be understood as an instanton-like configuration
from 2 pairs of D9, anti-D9 branes.
In the type I theory the gauge group
of a single D5-brane is Sp(1)=SU(2), and, as discussed in \cite{Witten}
(see also the review by Schwarz in ref. \cite{reva}), this seems to be
why one needs, in the construction, four
D9, anti-D9 pairs, characterised by a SO(4)$\times$ SO(4) gauge
symmetry. Indeed, in the review of Schwarz \cite{reva} it is argued
that one of the SU(2) of the SO(4) Chan-Paton group characterising say
the four anti-D9 branes, is used in the construction of the D5-brane
as an instanton configuration and the other one survives as its
gauge symmetry\footnote{This argument, which seems natural, is however
qualitative. Indeed, it is not, a priori, obvious to show
explicitely in this construction the connection between the remaining
SU(2) and the gauge group of the D5-brane.  
We thank the referee for a discussion on this point.}.

The D1-brane however is constructed from the same number of D9, anti-D9
pairs of branes in the type I and type IIB theories.
One may wonder how this is compatible
with the intermediate step in which a D1-brane is realised 
as a bound state
(D5, anti-D5) in type I \cite{Sen1},
whereas in the type IIB case one has
that D1=2 (D5, anti-D5) \footnote{This is implied by the structure of
the D5-brane worldvolume effective action, which contains a term:

\begin{equation}
\int_{R^{5+1}}{\rm Tr}\left[ C^{(2)}\wedge F\wedge F\right]\, .
\end{equation}

\noindent Again an SU(2) gauge structure is needed in order to 
have an integer instanton number.}. The presence of a different 
number of D5, anti-D5
branes is due to the fact that in the type I theory a single
D5-brane carries already an SU(2) gauge group, and therefore the D1-brane
can be realised as a bound state of a single pair D5, anti-D5
(see \cite{Sen1}). Therefore we can conclude that in terms of 9-branes:
D1=(D5, anti-D5)=8 (D9, anti-D9), using that in type I D5=4 (D9, anti-D9).

\subsection{The Heterotic case}

It was argued in \cite{Hull2} \cite{BEHHLvdS} that the Heterotic
string with gauge group SO(32) could be obtained as a nonperturbative
orientifold construction of the type IIB theory. This construction is
determined by S-duality: If the type I theory can be defined as an
orientifold of the type IIB theory by its worldsheet parity reversal
symmetry, its S-dual, i.e. the Heterotic SO(32) should be defined at
strong coupling as an orientifold of the type IIB theory by the
S-dual operation of worldsheet parity reversal. Perturbatively, this
operation coincides with the $(-1)^{F_L}$  symmetry of the
type IIB theory, where $F_L$ is the left-moving fermion number.
Modding out the type IIB theory by $(-1)^{F_L}$ gives
rise to the type IIA string \cite{Dab}, and this is determined 
by the twisted
sector that has to be added in order to restore modular invariance.
However one can consider adding anomaly cancelling 9-branes 
as in the type I
theory, and this is the way in which the Heterotic string 
can be produced. S-duality
determines that the background 9-branes must be NS9-branes, and also
that, nonperturbatively, $(-1)^{F_L}$ can be defined as the operation
reversing the orientation of the D-string. This worldsheet operation has
associated an orientifold fixed plane with -32 NS-NS charge, and this
charge is cancelled through the addition of 32 NS9-branes, with one
unit of NS-NS charge. Together they
reproduce the gauge sector of the Heterotic SO(32) supergravity
\cite{BEHHLvdS}, including
corrections in $\alpha^\prime$ already computed in the literature
\cite{Tseytlin}. 

The massless spectrum of the Heterotic
F-string can be described in terms of open D-strings with both ends 
on the F1, DD D-strings,
and with one end on the F1 and the other on an NS9-brane, 
DN D-strings, \cite{BEHHLvdS}.
The DD D-strings contribute with 8 scalars and 8 right-handed 
Majorana-Weyl spinors, whereas the DN D-strings contribute with 32
left-moving fermions. The analysis is completely analogous to that in
\cite{PW}, which shows that the BPS D-string of the type I theory
has the same worldsheet
structure than the SO(32) Heterotic string.
In that case open DD and DN F-strings provide the D-string with
the right massless
field content of the Heterotic.
In our case the open F-strings are replaced by open DD and DN D-strings,
which have however the same massless sector than the F1-branes. 
This description
of the Heterotic string arises at strong coupling. However,
given that the massless states are BPS, it can be as well extrapolated
to the weak coupling regime. In this limit the SO(32) charges tethered
to the F1 are pulled onto its worldsheet, since 
$\tau_{{\rm D1}}/\tau_{{\rm F1}}\sim 1/g_s\rightarrow\infty$ and the
D-strings collapse to a point, giving the usual SO(32) Heterotic
worldsheet currents.

So far the discussion has been focussed on the massless BPS states.
The Heterotic SO(32) theory contains  
as well perturbative massive
states in the spinorial representation of SO(32), which are non-BPS
but stable, given that they are the lightest ones transforming as
spinors of SO(32). Sen \cite{Sen1}, \cite{Sen2} 
showed that the correct way of 
describing these states at strong coupling is in terms of a weakly
coupled type I (D1, anti-D1) system. This configuration is unstable
due to the presence of a tachyonic mode in the open strings stretched
between the two branes, and moreover, since the two branes are
spinors under SO(32) \footnote{A BPS D1-brane carries the 
quantum numbers of the
spinorial representation of SO(32) \cite{Pol}.}
the bound state cannot transform as a spinor.
However, compactifying the D-strings and switching on a $Z_2$ Wilson
line the tachyonic mode can condense into a stable configuration
different from the vacuum, and transforming as a spinor of SO(32)
\cite{Sen1}. The mass of this configuration was also computed in 
\cite{Sen1} and shown to correspond to a D-particle $\sqrt{2}$ times
heavier than the BPS D-particle of the type IIA theory. In the convention
of the second reference in \cite{reva} this mass reads: 
$M=\frac{\sqrt{2}}{\sqrt{\alpha^\prime}}\frac{1}{g_I}$. 
At strong coupling one has, in the Heterotic side:

\begin{equation}
\label{masafu}
M=\frac{\sqrt{2}}{\sqrt{\alpha^\prime}}\sqrt{g_H}\, ,
\end{equation}

\noindent and this allows to determine how the mass of the perturbative
spinorial states in the Heterotic theory, given by 
$M=\frac{2}{\sqrt{\alpha^\prime}}$, gets renormalised as one increases
the Heterotic coupling constant. 

We can now take the following point of
view to describe the spinorial non-BPS states of the Heterotic theory
in the strong coupling regime.
A bound state D0=(D1, anti-D1) in the weakly coupled type I side 
predicts a bound state
(F1, anti-F1) in the strongly coupled Heterotic theory. 
S-duality determines that each of these
F1's transforms in the spinor representation of SO(32), since DN
open D-strings contribute with these quantum numbers, and that the whole
system is unstable due to the presence of a tachyonic mode in the open
D-strings stretched between the two F1's. S-duality determines as well
that if the system is compactified a state transforming as a spinor
under SO(32) should emerge after the condensation of the tachyonic mode.
The mass of this state is formally given by the same expression as in
the weakly coupled type I case. In the D0 = (D1, anti-D1) system the mass
is evaluated at the critical radius $R_c$ where the effective
mass square of all the tachyonic excitations becomes non-negative and the
two lightest ones have zero mass \cite{Sen1}.  This yields to  
$M_{\rm D0}=2(2\pi R_c)\tau_{\rm D1}$ where 
$ R_c=\sqrt{\frac{\alpha^\prime}{2}}$. Then it is argued \cite{Sen1} 
that this mass is actually independent of the radius of the compactification, 
providing us with the mass of the D-particle in the non-compact limit.
Now, in the (F1, anti-F1) Heterotic system
the corresponding strong coupling radius is 
given by ${\tilde R}_c = \sqrt{\frac{\alpha^\prime}{2}}\sqrt{g_H}$ and
the mass is
$M=2(2\pi {\tilde R}_c)\tau_{\rm F1} =
\sqrt{\frac{2}{\alpha^\prime}}\sqrt{g_H}$,
where we used $\tau_{\rm F1}=\frac{1}{2\pi\alpha^\prime}$. 
This provides an alternative interpretation of the result 
(\ref{masafu}) in terms of an (F1, anti-F1) bound state. 
This description at strong coupling is also natural from another point
of view. In the Heterotic theory the non-BPS states in the spinorial
representation of SO(32) correspond to unwrapped fundamental strings,
i.e. to strings not charged with respect to the NS-NS 2-form potential.
This charge cancellation is simply achieved by the F1, anti-F1
superposition. 

It is also interesting to point out that 
the Heterotic fundamental string can arise as a bound
state of a pair of NS5, anti-NS5 branes, since this is the S-dual 
configuration of the D1=(D5, anti-D5) bound state in type I.
This is implied by S-duality and can be read as
well directly from the NS5-brane effective action truncated
to a Heterotic background, in particular from the term \cite{EJL}:
$\int_{R^{5+1}}{\rm Tr}\left[ B^{(2)}\wedge {\tilde F}\wedge {\tilde F}
\right]$. 
In turn, the NS5-brane is obtained as a bound state of four NS9,
anti-NS9 pairs of branes. This is S-dual to the D5=4(D9, anti-D9)
configuration in type I, and can also be derived from the 
NS9-brane effective
action (\ref{NS9ac}) truncated to a Heterotic 
background\footnote{This implies that the Heterotic F-string can be
obtained as the bound state of eight NS9 brane-antibrane pairs
\cite{Witten}.
This can also be read from the NS9-brane effective action.}.
As we discussed in the previous section, from the SO(4) gauge
structure of four D9, anti-D9 pairs of branes only an SU(2) 
subgroup survives
after the instanton construction, and it remains as the gauge group
of the D5-brane. In a Polchinski-Witten type of analysis this
Yang-Mills multiplet with SU(2) gauge symmetry comes from open strings
with both ends on the D5-brane, whereas the D5-brane SO(32) currents
arise from DN open strings, i.e. from strings with one end on the
5-brane and the other on a D9-brane \cite{Witten2}.
Hull \cite{Hull2} argued that
the same massless modes should be present on the Heterotic 5-brane
worldvolume, in this case coming from open D-strings with DD and DN
boundary conditions, and with NS9-branes as the spacetime-filling
branes. Therefore, the
NS5, anti-NS5 pair of branes has an SU(2) $\times$ SO(32) gauge
structure, from which the SU(2)
is used in the instanton construction of the F-string and the SO(32)
group survives as its gauge structure. 

Finally, let us mention that S-duality
seems to imply that the S-dual of the non-BPS type I
D-instanton has the effect of breaking the O(32) gauge
group of the Heterotic at strong coupling\footnote{Because of its
duality with weakly coupled type I.} to its SO(32) gauge symmetry
group (see \cite{Witten}).

\section{Conclusions}

We have presented a description of the NS-NS p-branes of the type
IIB theory in terms of bound states of pairs of NS9, anti-NS9 
spacetime-filling branes. This description is determined by the
S-duality symmetry of the theory and it is also supported by the
worldvolume structure of the NS9-brane. From this construction
it is clear that the K-theory groups classifying the 
conserved NS-NS charges in type
IIB are again given by ${\tilde K}(S^{9-p})$, but now the 
spacetime-filling branes are NS9-branes.

Orientifolding this, strongly
coupled, type IIB theory by the S-dual of worldsheet parity reversal
we have provided a description of the Heterotic SO(32) branes as
topological solitons in systems of NS9, anti-NS9 pairs of 
branes truncated to a Heterotic background. The picture
that emerges is S-dual to the stable brane spectrum of type I
in terms of bound states of D9, anti-D9 branes, and gives an interesting
description of the strongly coupled non-BPS spinorial states of the
Heterotic SO(32) in terms of bound states (F1, anti-F1), as well as a
description of the F-string as a bound state (NS5, anti-NS5).

T-duality on the strongly coupled type IIB theory also gives 
interesting predictions in the type IIA side.
We have seen that the type IIB fundamental
string can be obtained as a bound state (NS5, anti-NS5) as:
F1=(D3, anti-D3)=2 (NS5, anti-NS5), where the tachyonic mode of a
D-string condenses. The T-dual picture describes a fundamental string
as a bound state of two NS5, anti-NS5 branes, where first the 
tachyonic mode of a D2-brane condenses and then the tachyonic
mode of a different brane, a D0-brane:
F1=(D2, anti-D2)=2 (NS5, anti-NS5)\footnote{We can alternatively
consider the process: F1=(D4, anti-D4)=2(NS5, anti-NS5). Now first
a D0-brane condenses and then a D2-brane. These intermediate processes
have been discussed in \cite{Yi,LY}.}. {}From the worldvolume
effective action of the type IIA NS5-brane this is described by the
term:
$\int_{R^{5+1}}B^{(2)}\wedge da^{(2)}\wedge dc^{(0)}$,
where $a^{(2)}$ and $c^{(0)}$ denote, respectively, the 2-form and
scalar worldvolume fields of the NS5A-brane (see \cite{BLO}).

The T-dual picture of our description of type IIB branes from
NS9, anti-NS9 systems gives a realisation
of type IIA branes as bound states of NS9A-branes,
which are T-dual to the NS9-branes of the type IIB theory.
This 9-brane is predicted by the analysis of the type IIA spacetime
supersymmetry algebra \cite{Hull1}, but contains a Killing direction in
its worldvolume, something that is determined both from the T-duality
transformation and from its eleven dimensional description as an
M9-brane (see \cite{BEHHLvdS}), so that it is really a domain-wall type of
solution and not a spacetime-filling brane. Therefore, it seems that 
if one wants to have brane descent constructions which preserve ten
dimensional Lorentz invariance one is constrained
to use the non-BPS but spacetime-filling D9-branes introduced by
Ho\u{r}ava in \cite{Horava}.

Finally, it would also be interesting to analyse similar configurations
to the ones studied in this paper
in the Heterotic $E_8\times E_8$ theory. T-duality on the strongly
coupled Heterotic SO(32) theory should provide interesting configurations
in the strongly coupled Heterotic $E_8\times E_8$ side, in its 
description as a nonperturbative orientifold of the type IIA
theory \cite{BEHHLvdS}. We hope to report progress in this direction in
the near future.

\subsection*{Acknowledgements}

L.~H. would like to acknowledge the support of the European Commission 
TMR programme grant ERBFMBICT-98-2872, and the partial support of PPARC
from the grant PPA/G/S/1998/00613.

\end{document}